\documentstyle[12pt]{article}
\input epsf
\begin{document}
%%%%%%%%%%%%%%%%%%%%%%%%%%%%%%%%%%%%%%%%%%%%%%%%%%%%%%%%%%%%%%%%%%%%%%
%%%%%%%%%%%%%%%%%%%%%%%%%%%%%%%%%%%%%%%%%%%%%%%%%%%%%%%%%%%%%%%%%%%%%%
\thispagestyle{empty}
\begin{titlepage}
%\maketitle
%%%%%%%%%%%%%%%%%%%%%%%%%%%%%%%%%%%%%%%%%%%%%%%%%%%%%%%%%%%%%%%%%%%%%%
%%%%%%%%%%%%%%%%%%%%%%%%%% Preprint Number %%%%%%%%%%%%%%%%%%%%%%%%%%%
%%%%%%%%%%%%%%%%%%%%%%%%%%%%%%%%%%%%%%%%%%%%%%%%%%%%%%%%%%%%%%%%%%%%%%
\begin{flushright}
        {\normalsize
NIIG-DP-99-04\\
%hep-ph/yymmddd\\
December, 1999   }\\
\end{flushright}
%%%%%%%%%%%%%%%%%%%%%%%%%%%%%%%%%%%%%%%%%%%%%%%%%%%%%%%%%%%%%%%%%%%%%%
\vspace{0.5cm}
\begin{center}
\Large\bf
{\Huge Four-Dimensional Planck Scale is Not Universal
in Fifth Dimension in Randall-Sundrum  Scenario }
\end{center}
\vspace{1cm}

\setcounter{footnote}{0}
\begin{center}
{\large
Takaaki~{\sc Ozeki}\footnote{ozeki@muse.hep.sc.niigata-u.ac.jp}
and
Noriyuki~{\sc Shimoyama}\footnote{simoyama@muse.hep.sc.niigata-u.ac.jp}
}\\
\vspace{0.5cm}
{\it
Department of Physics, Niigata University, %\\
Niigata 950-2182, Japan}\\
\vspace{0.5cm}
\end{center}
\vskip 0.5cm
\begin{abstract}
It has recently been proposed that the hierarchy problem can be solved 
 by considering the warped fifth dimension compactified on
 $S^{1}/Z_{2}$.
Many studies in the context have assumed a particular choice for 
an integration constant $\sigma_{0}$ that appears when one solves 
the five-dimensional Einstein equation. Since $\sigma_{0}$ is not
determined by the boundary 
condition of the five-dimensional theory, $\sigma_{0}$ may be regarded as 
a gauge degree of freedom in a sense.
To this time, all indications are that the four-dimensional Planck mass 
depends on $\sigma_{0}$. 
In this paper, we carefully investigate the properties of the geometry in
the Randall-Sundrum model,
and consider in which loction $y$ the four-dimensional Planck mass is 
measured.
As a result, we find a  $\sigma_{0}$-independent relation between
the four-dimensional Planck mass $M_{\rm Pl}$ and the five-dimensional
fundamental mass scale $M$, and remarkably enough, we can take $M$ 
to the TeV region when we consider models with the Standard Model confined 
on a distant brane.
We also confirm that the physical masses on the distant brane do not 
depend on $\sigma_{0}$ by considering a bulk scalar field as an illustrative 
example. 
The resulting mass scale of the Kaluza-Klein modes is on the order of $M$. 
\end{abstract}
\end{titlepage}
\setcounter{footnote}{0}
\date{\today}

\newpage
\section{Introduction}
 The vast gap between the electroweak  scale and the  Planck scale, 
known as the `hierarchy problem', remains as a major mystery in particle 
physics.
 Recently, it has been suggested that large compactified extra dimensions
may provide a solution to the hierarchy problem.\cite{Arkani}
The relation between the four-dimensional Planck scale $M_{\rm Pl}$ and 
the higher dimensional scale $M$ is given by 
$M^{2}_{\rm Pl}=M^{n+2} V_{n}$,  
where $V_{n}$ is the volume of the extra  compactified dimensions.
If  $V_{n}$ is large enough, $M$ can be on the order of several TeV.
Unfortunately, this scenario alone  does not completely solve the problem. 
The original hierarchy can be translated into another hierarchy between 
$M$ and the compactification scale $r^{-1}_{c}=V^{-1/n}_{n}$ .
 
In Ref.~\cite{Randall1}, Randall and Sundrum (RS) proposed an alternative 
scenario based on an extra dimension compactified on $S^{1} / Z_{2}$ with 
three-branes located at two boundaries.
Standard Model fields are assumed to be confined on a distant three-brane,
while gravitons propagate in the five-dimensional bulk. 
The background metric takes the form
\begin{eqnarray}
  ds^{2} =   G_{MN}dx^{M}dx^{N} 
         =    e^{-2\sigma(y)}\eta_{\mu\nu}dx^{\mu}dx^{\nu} + dy^{2},
\end{eqnarray}
where $x^{M}=(x^{\mu}, y)$ is the coordinate of the five-dimensional 
spacetime. 
By the solving the five-dimensional Einstein equation, the function 
$\sigma(y)$ is found.
\begin{eqnarray}
\sigma(y)=k|y|+\sigma_{0},
\end{eqnarray}
where $k$ is the curvature scale related to the five-dimensional 
cosmological constant $\Lambda$. 
It was then shown that the effective Planck mass in four dimensions
is given by the formula
\begin{eqnarray}
  M^{2}_{\rm Pl}
          &=& \frac{M^{3}}{k}(1-e^{-2\pi kr_{c}})e^{-2\sigma_{0}},
\label{eqn:pl}
\end{eqnarray}
where  $r_{c}$ is the radius of the fifth dimension.

Two points are to be noted here. The first point is that according to the 
above formula (\ref{eqn:pl}), the four-dimensional Planck mass 
is `universal'; it takes the same value on both branes located at different
boundaries. 
On the other hand, it was argued that mass scales on the distant brane
located at $y=\pi r_{c}$ are rescaled by the warp factor $e^{-\pi k r_{c}}$ .
It is this difference that generates the hierarchy 
\begin{eqnarray}
  \frac{M_{W}}{M_{\rm Pl}} \sim  \frac{M e^{-\pi k r_{c}}}{M}
  \ll 1 .
\end{eqnarray}
The second point to be noted is that the formula (\ref{eqn:pl})
apparently contains the integration constant $\sigma_{0}$, which is 
left undetermined by the boundary condition of the five-dimensional
theory.  Moreover, a certain kind of symmetry under the exchange of two
boundaries is not manifest in this formula (\ref{eqn:pl}), 
as is pointed out in Ref.\cite{Dienes}.

In this paper, we argue that these two points are intimately related. 
We carefully discuss the induced metric and the
brane coordinates,   
and point out that the value of the four-dimensional Planck mass 
differs for hidden and visible brane; that is,it is  non-universal.
As a result, we find that the four-dimensional Planck mass does not depend 
on $\sigma_{0}$, and exchanging-symmetry is manifest. 

This paper is organized as follows. 
After a review of the RS model in \S 2, we present our formula for
the Planck mass in \S 3.
In \S 4, we apply our prescription for a bulk scalar field as an
example and examine the masses of its Kaluza-Klein (KK) modes.\cite{Wise}
We show that these masses are also independent of $\sigma_{0}$.
As a result, in the $k>0$ scenario, we can adjust $M$ to the TeV region 
since the Planck mass as the gravitational coupling constant on the
distant brane should be set to  $10^{18}$ GeV, 
and we find
that the mass of the KK modes is on the order of $M$.   
Section 5 is devoted to conclusion and discussion.

\section{RS model}
First, we review the derivation of the four-dimensional Planck mass 
within the scenario of Ref.\cite{Randall1}.

The background metric of the model takes the form
\begin{eqnarray}
  ds^{2} =   G_{MN}dx^{M}dx^{N} 
         =    e^{-2\sigma(y)}\eta_{\mu\nu}dx^{\mu}dx^{\nu} + dy^{2},
\label{eqn:metric}
\end{eqnarray}
where $x^{M}=(x^{\mu}, y)$ is the coordinate of the five-dimensional 
spacetime, and the fifth dimension is compactified on $S^{1}/Z_{2}$ with 
radius $r_{c}$. 
The fundamental region of the fifth dimension is given by 
$0\leq y \leq\pi r_{c}$.
A set of three-branes is located at each fixed point $y=y_{i}$ of 
$S^{1}/Z_{2}$. 
The brane at $y_{0}= 0$ is called a `hidden' brane, and the
brane at $y_{1}= \pi r_{c}$ is called `visible'. 
Here and hereafter, we use the subscripts $i=0,1$ for quantities 
at $y=0$, $\pi r_{c}$, respectively.

The action is
\begin{eqnarray}
          S  &=& S_{\rm gravity}+ S_{\rm brane}  ,\nonumber\\
 S_{\rm gravity} &=& \int d^{4}x \int_{-\pi r_{c}}^{\pi r_{c}} dy \sqrt{-G} 
                     \left\{ -\Lambda + 2M^{3}R    \right\} ,\nonumber\\
  S_{\rm brane}  &=& \sum_{i=0,1}\int d^{4}x \sqrt{-g^{(i)}} \left\{ 
                           {\cal L}_{(i)}-V_{(i)}      \right\},
\end{eqnarray}
where $g^{(i)}_{\mu\nu}(x)=G_{\mu\nu}~(x,y=y_{i})$ is the induced
 metric on the ${i}$-th brane, and the brane tensions $V_{(i)}$ are 
subtracted from the three-brane Lagrangians.

With the metric (\ref{eqn:metric}), the five-dimensional Einstein
equation reduces to two differential equations for $\sigma(y)$ (using 
$\sigma'= \partial_{y}\sigma$) : 
\begin{eqnarray}
\label{eqn:prime}
(\sigma'(y))^{2}  &=& \frac{-\Lambda}{24M^{3}},\\ 
\sigma''(y)        &=& \sum_{i=0,1}\frac{V_{(i)}}{12M^{3}}\delta( y - y_{i}). 
\label{eqn:doubleprime}
\end{eqnarray}
The solution to (\ref{eqn:prime}) is given by
\begin{eqnarray}
  \sigma(y) = k|y| + \sigma_{0},
  \label{eqn:sigma}
\end{eqnarray}
with
\begin{eqnarray}
  k = \pm \sqrt{\frac{-\Lambda}{24M^{3}}}.  
\end{eqnarray}
Here, $\sigma_{0}$ is the integration constant which is not determined
 by the  boundary condition of the five-dimensional theory.\footnote
{The integration constant $\sigma_{0}$ might be
determined by the fundamental theory in higher dimensions.\cite{Verlinde}}
Conventionally, this integration constant $\sigma_{0}$ is omitted by
saying that it just amounts  to an overall constant rescaling of 
$x^{\mu}$. 
For any value of $\sigma_0$,
the consistency of the solution (\ref{eqn:sigma}) with the second
equation (\ref{eqn:doubleprime}) requires that
the brane tensions and bulk cosmological 
constant are related by 
\begin{eqnarray}
  k = \frac{V_{(0)}}{24M^{3}} = -\frac{V_{(1)}}{24M^{3}}.
\label{eqn:fine}
\end{eqnarray}
Following to Ref.\cite{Randall1}, we consider a fluctuation
$h_{\mu\nu}$ of the Minkowski spacetime metric $\eta_{\mu\nu}$, and replace 
$\eta_{\mu\nu}$  by a four-dimensional metric 
$\overline{g}_{\mu\nu}=\eta_{\mu\nu}+h_{\mu\nu}$. 
Then we have
\begin{eqnarray}
 S_{\rm eff} &=&  \int d^{4}x \int_{-\pi r_{c}}^{\pi r_{c}} dy\,  2M^{3}
            e^{-2\sigma(y)} \sqrt{-\overline{g}}R(\overline{g})+ \cdots
                                                       \nonumber \\
         &\equiv&  2 M^{2}_{\rm Pl}\int d^{4}x 
            \sqrt{-\overline{g}}R(\overline{g})+ \cdots \ ,
\end{eqnarray}
where $R(\overline{g})$ denotes the four-dimensional scalar curvature
constructed from $\overline{g}_{\mu\nu}$.
This gives the formula (\ref{eqn:pl}) for the `universal' four-dimensional 
Planck mass,\footnote
{We use the normalizaton of Ref.\cite{Randall1}.}
\begin{eqnarray}
  M^{2}_{\rm Pl}
 = \frac{M^{3}}{k}(1-e^{-2\pi kr_{c}})e^{-2\sigma_{0}} .
 \label{eqn:planck}
\end{eqnarray}
We stress that this expression for the Planck mass depends on the integration
constant $\sigma_{0}$.
With the choice $\sigma_{0}=0$, as in Ref.\cite{Randall1},
the above expression reduces to 
\begin{eqnarray}
 M^{2}_{\rm Pl}
 = \frac{M^{3}}{k}(1-e^{-2\pi kr_{c}}) .
  \label{eqn:RSplanck}
\end{eqnarray}
With this choice, one is forced to take $M$ to be of the order of 
$M_{\rm Pl}\sim 10^{18}$ GeV when considering  $k>0$.

As pointed out in Ref.\cite{Dienes}, we are  free 
to choose the $y$-independent constant $\sigma_{0}$ 
in Eq.~(\ref{eqn:sigma}).
The particular choice $\sigma_0=-\pi kr_{c}/2$ was made so as to meet 
the requirement that the expression for $M_{\rm Pl}$ is manifestly 
invariant with the respect to the change $k \rightarrow -k$, which amounts to
exchanging the role of the two boundaries.
Then, the four-dimensional Planck mass can be written 
\begin{eqnarray}
 M^{2}_{\rm Pl}= \frac{2M^{3}}{k}\sinh(2\pi kr_{c}) .
  \label{eqn:DDGplank}
\end{eqnarray}

As was noted in Ref.\cite{Dienes},
however, it is almost certainly true that a change 
of $\sigma_{0}$ has no net physical effect  and must not change 
the values of four-dimensional physical quantities. 
Therefore the physical quantities
should have the above exchanging-symmetry {\em without choosing $\sigma_0$}. 
In other words, physical quantities, including Planck mass, should have
the following two properties.
First, they are independent of $\sigma_0$. 
Second, they are not affected by the above brane-exchanging.

In the next section, we present a prescription that
naturally realizes these two properties.

\section{Four-dimensional effective Planck mass}

As stated above, the choice of $\sigma_{0}$ has no net physical effect, 
 and it must not change the values of physical quantities. 
That is, all the physical quantities, including the Planck mass,  must be 
independent of $\sigma_{0}$. 
In this sense $\sigma_{0}$ may be regarded as a gauge degree of freedom.
In particular, the $\sigma_{0}$-independence of the four-dimensional
Planck mass may be understood by the following argument. 
Observe that $\sigma_{0}$ determines the ratio of the length  
scales of the fifth-dimensional direction and four-dimensional
direction at $y=0$. 
Therefore, after we have integrated over the full 
fifth dimension when calculating $M_{\rm Pl}$, the freedom of this
ratio will be invisible in the effective theory.  

To find the four-dimensional Planck mass more carefully,
it is important to make it clear in which location $y$ the 
four-dimensional Planck mass is measured. 
To this end, we need to reconsider the choice of the brane coordinate 
and the induced metric.

We first recall the general situation. 
When the $i$-th brane with the brane coordinate $\xi^{\mu}_{i}$ is 
embedded into five-dimensional spacetime by $x^{\mu}=x^{\mu}(\xi_{i})$ 
and $y=y_{i}$, 
the induced metric on it is given by 
\begin{eqnarray}
g^{(i)}_{\mu\nu}
\left(\xi_{i}\right)
=
\frac{\partial x^{M}}{\partial\xi_{i}^{\mu}}
\frac{\partial x^{N}}{\partial\xi_{i}^{\nu}}
G_{MN}
\left(
x=x
\left(
\xi_{i}
\right)
,
y=y_{i}
\right).
\label{eqn:inducedmetric}
\end{eqnarray} 
In the discussion given in \S 2,
the implicit choice $x^{\mu}=\xi^{\mu}_{i}$ was made so that 
$g^{(i)}_{\mu\nu}=G_{\mu\nu}(y=y_{i})$.

When the fine-tuning conditions (\ref{eqn:fine}) are satisfied,
the branes are flat and the induced metrics generally take the form 
$g_{\mu\nu}^{(i)}= e^{-2\alpha}\eta_{\mu\nu}$, with the $x^{\mu}$-independent 
constant $\alpha$.
In view of Eq.~(\ref{eqn:inducedmetric}), the corresponding brane
coordinates are uniquely determined by 
$\xi^{\mu}_{i}=e^{\alpha-\sigma_{i}}x^{\mu}$ 
(up to a Poincar\'{e} transformation).
Therefore, when discussing a four-dimensional effective theory on the 
brane, one should use the correct set of the induced metric and brane 
coordinate as 
\begin{eqnarray}
\left(g_{\mu\nu}^{(i)}=e^{-2\alpha}\eta_{\mu\nu},
~\xi_{i}^{\mu}=e^{\alpha-\sigma_{i}}x^{\mu}
\right).
\label{eqn:general}
\end{eqnarray}
This is the point of our treatment. With this correct set, 
we can determine the relation of the five-dimensional scale M 
and the four-dimensional Planck scale $M_{{\rm Pl} (i)}$ on the $i$-th
brane by integrating over the fifth dimension as 
\begin{eqnarray}
S_{eff}
&=&
2M^{3}\int d^{4}x\int^{\pi r_{c}}_{-\pi r_{c}}dy\,
e^{-2\sigma\left(y\right)}
\sqrt{-{\overline g}}R\left({\overline g};x\right)
+
\cdots
\nonumber\\
&=&
2M_{{\rm Pl}(i)}^2
\int d^{4}\xi_{i}\,
\sqrt{-{\overline g}^{(i)}}
R
\left(
{\overline g}^{(i)};\xi_{i}
\right)
+
\cdots,
\end{eqnarray}
where $R\left({\overline g}^{(i)},\xi_{i}\right)$ is the
four-dimensional scalar curvature constructed from 
the induced metric ${\overline g}_{\mu\nu}^{(i)}$ and the coordinate 
$\xi_{i}^{\mu}$.
We note that, following Ref.\cite{Randall1}, we define $M_{{\rm Pl}}$ as 
a coefficient of the Einstein-Hilbert (EH) term. 
This is a proper definition of graviton self-couplings contained in 
the EH term.
Alternatively, we can determine $M_{{\rm Pl}}$ from the graviton coupling to 
the matter stress tensor. 
We confirmed that both methods yield the same result, given below.

We now determine the relation between $M_{{\rm Pl}}$ and
the five-dimensional quantities  $M,~k and r_{c}$ by using the set 
(\ref{eqn:general}).
For definiteness, let us choose $\alpha=0$ so that the induced
metric is precisely Minkowskian (as is usual in field theory in flat
space-time).
This means that we choose 
($\eta_{\mu\nu},\xi^{\mu}_{i}=e^{-\sigma_{i}}x^{\mu}$)
as the induced metric and brane coordinate. 
With this choice, we change the integration variables to
$\xi^{\mu}_{i}=e^{-\sigma_{i}}x^{\mu}$ and contract the indices
with $g^{(i)}_{\mu\nu}=\eta_{\mu\nu}$.
We then obtain 
\begin{eqnarray}
S_{eff}  
& = & 
\int d^{4}\xi_{i} 
\int_{-\pi r_{c}}^{\pi r_{c}} 
dy ~2M^{3} e^{-2(\sigma(y)-\sigma_{i})}
\sqrt{-\overline{g}^{(i)}}R(\overline{g}^{(i)};\xi_{i})                   
+\cdots 
\nonumber\\     
&\equiv& 2 M^{2}_{{\rm Pl}(i)} 
\int d^{4}\xi_{i}
\sqrt{-\overline{g}^{(i)}}R(\overline{g}^{(i)}; \xi_{i})
+\cdots,       
\end{eqnarray}     
where $\sigma_{i}=\sigma(y_{i})$.
It follows that
\begin{eqnarray}
 M^{2}_{{\rm Pl}(i)}
    &=& \int_{-\pi r_{c}}^{\pi r_{c}} dy M^{3} e^{-2(\sigma(y)-\sigma_{i})} 
        \nonumber \\
    &=& M^{3}e^{2(\sigma_{i}-\sigma_{0})}
        \int_{-\pi r_{c}}^{\pi r_{c}}dy e^{-2k|y|} 
        \nonumber\\
    &=& \frac{M^{3}}{k}(1 - e^{-2\pi kr_{c}})e^{2(\sigma_{i}-\sigma_{0})}.
\label{eqn:gauge}
\end{eqnarray}               
With (\ref{eqn:gauge}), it is clear that $M_{{\rm Pl}(i)}$ is  
independent of $\sigma_{0}$. 
Explicitly, we find 
\begin{eqnarray}
 M^{2}_{\rm Pl(0)} = \frac{M^{3}}{k}(1 - e^{-2\pi kr_{c}})
 \label{eqn:OSplanck(0)}  
\end{eqnarray}
on the brane at $y=0$ and
\begin{eqnarray}
M^{2}_{\rm Pl(1)} = \frac{M^{3}}{k}(e^{2\pi kr_{c}} - 1 )
\label{eqn:OSplanck(1)}
\end{eqnarray}
on the brane $y=\pi r_{c}$.
Note that these expressions are transformed into each other 
by exchanging $k$ with $-k$.
Thus our results naturally possess the two properties stated in the previous 
section.

We note that our expression (\ref{eqn:OSplanck(0)}) for the brane at $y=0$
coincides with Eq.~(\ref{eqn:RSplanck}), which is derived by simply
neglecting $\sigma_{0}$. 
Therefore in this case, we have explicitly confirmed the naive
expectation that $\sigma_{0}$ may be absorbed by the rescaling of 
$x^{\mu}$,
since our expression (\ref{eqn:OSplanck(0)}) takes account of 
$\sigma_{0}$ by using the correct set 
($\eta_{\mu\nu}, \xi^{\mu}_{i=0}=e^{-\sigma_{0}}x^{\mu}$) of the 
induced metric and brane coordinate.

The same is not true for the expression (\ref{eqn:OSplanck(1)}), 
however.
When we consider the scenario of Ref.\cite{Randall1}, in which Standard
Model fields are assumed to be confined on the brane at $y=\pi r_{c}$
with a negative tension,
the naive expectation is no longer correct, and the original expression 
(\ref{eqn:RSplanck}) should be modified to our (\ref{eqn:OSplanck(1)}).
The origin of the discrepancy can be understood as follows.
If one tries to absorb $\sigma_{0}$ by the rescaling 
$\xi^{\mu}_{i=1}=e^{-\sigma_{0}}x^{\mu}$ as in the $y=0$ case, 
the induced metric $\eta_{\mu\nu}$ cannot be used, 
since ($\eta_{\mu\nu}, \xi^{\mu}_{i=1}=e^{-\sigma_{0}}x^{\mu}$) is not
the correct set of the induced metric and brane coordinate.
The correct set is ($\eta_{\mu\nu},
\xi^{\mu}_{i=1}=e^{-\sigma_{1}}x^{\mu}$).
Usihg this correct set, one finds that the Planck scale at 
$y=\pi r_{c}$ is given by our formula (\ref{eqn:OSplanck(1)}). 

The most important aspect of the RS model is that it gives rise to 
a localized graviton field.\cite{Randall2}
Our results (\ref{eqn:OSplanck(0)}) and (\ref{eqn:OSplanck(1)}) 
can naturally be understood from this fact;  
the small Planck scale $M_{\rm{Pl}(0)}$
arises because of the localized graviton in the fifth dimension near 
the brane of positive tension, while the large Planck scale 
$M_{\rm{Pl}(1)}$ arises 
because of the small overlap of the graviton  with the brane of 
negative tension.\footnote
{To be precise, $M_{\rm P1(1)}$ should be identified as the Planck scale
in models with the Standard Model confined on the brane at $y=\pi r_{c}$,
and $M_{\rm Pl(0)}$ should be identified as that in models with the SM
confined at $y=0$.
In any case, we have only one Planck scale in a given model.}

A striking feature of our results (\ref{eqn:OSplanck(0)}) and 
(\ref{eqn:OSplanck(1)}) is that the relative size of four and
five-dimensional Planck scales crucially depends on the location at
which $M_{\rm Pl}$ is measured.
In the model in which Standard Model fields are confined on the positive 
tension brane at $y=0$, as in Ref.\cite{Randall2},
%In the scenario Ref.\cite{Randall2}, 
we have the relation (\ref{eqn:OSplanck(0)}), 
from which $M$ is of the same order as $M_{{\rm Pl}(0)} \sim
10^{19}$ GeV, that is, $M \sim M_{{\rm Pl} (0)}$.
This conclusion is the same as that in the original proposal, of course.
In the model in which Standard Model fields are confined on 
the negative tension brane at $y=\pi r_{c}$, as in Ref.\cite{Randall1},  
%In the scenario of Ref.\cite{Randall1}, 
however, we now have our modified 
relation (\ref{eqn:OSplanck(0)}), which gives 
\begin{eqnarray}
M^{2}_{\rm{Pl}(1)} \approx \frac{M^{3}}{k}e^{2\pi kr_{c}}.
\label{eqn:limit}
\end{eqnarray}
We see that the fundamental mass scale $M$ becomes much smaller than 
the Planck scale, unlike in the original proposal.\cite{Randall1}
As a result, it is perfectly possible that the fundamental scale $M$
lies in the TeV region.
We need to use the expressions (\ref{eqn:OSplanck(0)}) and 
(\ref{eqn:OSplanck(1)}) of the four-dimensional Planck mass 
properly, depending on the model employed.

\section{Physical mass scale}
In order to check from another viewpoint that the physical quantities are 
independent of $\sigma_{0}$, we consider a massless bulk scalar field as 
an illustrative example, and examine the masses of the KK modes.
Extensions to the case of a massive bulk scalar field and a bulk gauge field are
straightforward.

The action is given by
\begin{eqnarray}
 S_{\rm scalar} &=& -\frac{1}{2}\int d^{5}x \sqrt{-G}G^{MN}
     \partial_{M}\Phi\partial_{N}\Phi     \nonumber \\
            &=&~~\frac{1}{2}\int d^{5}x e^{-2\sigma}\Phi
     (\Box +e^{2\sigma}\partial_{y}e^{-4\sigma}\partial_{y})\Phi,
\end{eqnarray}
where  $\Box \equiv {\eta}^{\mu\nu} \partial_{\mu} \partial_{\nu}$.
The KK mode expansion is 
\begin{eqnarray}
 \Phi(x, y) =\sum_{n\geq 0}\varphi_{n}(x)\chi_{n}(y),
\end{eqnarray}
where the mode functions $\chi_{n}(y)$ are chosen to satisfy
\begin{eqnarray}
 \frac{d}{dy}\left( e^{-4\sigma}\frac{d}{dy}\right)\chi_{n}(y)
   = -M^{2}_{n}e^{-2\sigma}\chi_{n}(y)
\label{eqn:bessel}
\end{eqnarray}
with mass eigenvalues $M_n$.
The solutions to this equation are related to Bessel functions $J_{2}$
and $Y_{2}$ of order two.\cite{Wise} 
We should be careful with Eq.~(\ref{eqn:bessel}) however, since it still
contains  $\sigma_{0}$.
As the orthnormality condition for $\chi_{n}$,
let us take
\begin{eqnarray} 
\int_{-\pi r_{c}}^{\pi r_{c}} dy e^{-2\sigma}\chi_{m}\chi_{n}=
\delta_{mn}e^{-2\sigma_{i}}.
\end{eqnarray}
Then we have 
\begin{eqnarray}
S_{\rm scalar}&=&  \frac{1}{2}\sum_{n\geq 0}
\int d^{4}x~e^{-2\sigma_{i}}
\varphi_{n}(\Box -M^{2}_{n})\varphi_{n}\,.
\end{eqnarray}

From  our point of view, the action should be described by using the 
$i$-th brane coordinate $x^{\mu}_{i}=e^{-\sigma_{i}}x^{\mu}$.
Then, the four-dimensional volume element $d^{4}x$ and 
the differential operator $\Box$ are replaced by
 $d^{4}x_{i}=d^{4}xe^{-4\sigma_{i}}$ and $\Box_{i}=\Box e^{2\sigma_{i}}$, 
respectively:
\begin{eqnarray}
S_{\rm scalar}
&=&
\frac{1}{2}\sum_{n\geq 0}
\int d^{4}x_{i}\,\varphi_{n}^{\prime}
\left(\Box_{i}-e^{2\sigma_{i}}M_{n}^{2}
\right)
\varphi_{n}^{\prime}\,.
\end{eqnarray}
We find that canonically-normalized fields in four dimensions
are $\varphi_{n}^{\prime}(x_{i})=\varphi_{n}(x)$, 
and the physical masses of the KK modes are given by
\begin{eqnarray}
{M_{n(i)}^{\prime 2}}\equiv e^{2\sigma_{i}}M_{n(i)}^2\,.
\label{eqn:kkmass}
\end{eqnarray}

To clarify the physical meaning of (\ref{eqn:kkmass}),
let us rewrite (\ref{eqn:bessel}) as
\begin{eqnarray}
\frac{d}{dy}
\left(
e^{-4ky}\frac{d}{dy}
\right)
\chi_{n}
=
-\left(
M_{n}^2e^{2\sigma_{0}}
\right)e^{-2ky}
\chi_{n}
\equiv
-m_{n}^{2}e^{-2ky}
\chi_{n}\,.
\label{eqn:barebessel}
\end{eqnarray}
With $m_{n}^{2}$ regarded as an eigenvalue, this equation is independent 
of $\sigma_{0}$.
This implies that $m_{n}^{2}$ depends on the parameters $k$ and $r_{c}$,
but not on $\sigma_{0}$;
$m_{n}^{2}=m_{n}^2(k,r_{c})$.
Therefore the $\sigma_{0}$ dependence cancels out in
Eq.~(\ref{eqn:kkmass}) as
\begin{eqnarray}
M_{n(i)}^{\prime}
=
M_{n}e^{\sigma_i}
=
m_{n}e^{\left(\sigma_{i}-\sigma_{0}\right)}\,.
\end{eqnarray}

The mass spectrum of the KK modes is  determined by
 the  boundary condition at $y=\pi r_{c}$, 
\ $\left(d/dy\right)\chi_{n}\left(y=\pi r_{c}\right)=0$. 
In particular, we are interested in the mass scale $M_{n(1)}^{\prime}$,
 as measured on the visible brane.
To this end, note that Eq.~(\ref{eqn:barebessel}) is precisely the same 
 equation that treated in Ref.\cite{Wise},
where it was shown that
\begin{eqnarray}
m_{n}\sim ke^{-\pi kr_{c}}=ke^{-\left(\sigma_{1}-\sigma_{0}\right)}
.
\end{eqnarray}
Therefore the mass scale of the KK modes is estimated as 
\begin{eqnarray}
M_{n(1)}^{\prime}
\sim
k
\sim
M
\ll M_{\rm Pl(1)}
\,.
\end{eqnarray}
We thus confirm that the mass scale of the KK modes is significantly
smaller than the Planck scale $\sim 10^{18}$ GeV.
We remark, however, that our interpretations is  quite different from 
the usual one,
\begin{eqnarray}
m_{n}
\ll
k
\sim
M
\sim
M_{\rm Pl}
\,. 
\end{eqnarray}

\section{Conclusions}
The most important point in our treatment is to make the brane coordinate
transformation,
 by which the induced metric on each brane becomes the Minkowski 
metric $\eta_{\mu\nu}$.
When measured with such brane coordinates, the four-dimensional Planck 
mass becomes different on each brane.
We found that the four-dimensional Planck mass, when
calculated by this procedure, does not depend on the integration constant
 $\sigma_{0}$.
We showed that the masses of the KK modes are also independent  
of $\sigma_0$ for a massless bulk scalar.
This holds for any  kinds of fields.
Moreover, the exchanging-symmetry is manifest in our expression 
for the Planck masses.
On the brane with negative tension,
the Planck mass is always much larger than on the brane with positive
tension.
This fact is interpreted as reflecting the smallness of the gravitational 
coupling constant, that is, the small overlap of the graviton with the 
brane with negative tension.
When we identify the brane of  negative tension
with the visible brane,
the fundamental scale $M$ can be significantly smaller than the
Planck mass,
and the masses of the KK modes are of the same order as $M$ in the
massless bulk scalar.
We can summarize the above statement as follows.
When we estimate the values of  physical quantities,
we must multiply by the warp factor corresponding to their mass dimension 
to the values in the bulk.
This rule is not exceptional to the Planck mass. 

The most striking result is that the fundamental scale $M$ in the
five-dimensional theory can be significantly smaller than that was 
supposed so far. 
For instance, it can lie in the TeV region.
Given this, it will be interesting to find direct evidence for the
extra dimension in the Randall-Sundrum-type scenario.
High-energy accelerator experiments in the near future might directly
prove the existence of the extra dimension.

\section*{Acknowledgements}

We are grateful to H.~Nakano, A.~Kageyama and T.~Hirayama 
for useful discussions and suggestions. We also thank K.-I.~Izawa for
helpful comments.


\noindent  

\begin{thebibliography}{99}
\bibitem{Arkani}
N.~Arkani-Hamed, S.~Dimopoulos and G.~Dvali,
``{\it The Hierarchy Problem and New Dimensions at a Millimeter}'', 
Phys. Lett. {\bf B429} (1998) 263, hep-ph/9803315.\\
I.~Antoniadis, N.~Arkani-Hamed, S.~Dimopoulos and G.~Dvali,
``{\it New Dimensions at a Millimeter to a Fermi and Superstrings at a TeV}''
Phys.~Lett.\ {\bf B436} (1998), 257, hep-ph/9804398.

\bibitem{Randall1}
L.~Randall and R.~Sundrum,
``{\it A Large Mass Hierarchy from a Small Extra Dimension}'', 
Phys. Rev. Lett. {\bf 83} (1999) 3370, hep-th/9905221.

\bibitem{Randall2}
L.~Randall and R.~Sundrum,
``{\it An Alternative to Compactification}'', 
Phys. Rev. Lett. {\bf 83} (1999) 4690, hep-th/9906064.


\bibitem{Dienes}
K.R.~Dienes, E.~Dudas and T.~Gherghetta,
``{\it Anomaly-Induced Gauge Unification and Brane/Bulk Coupling in 
Gravity-Localized Theories}'',
hep-ph/9908530.

\bibitem{Wise}
W.D.~Goldberger and  M.~Wise,
``{\it Bulk Field in the Randall-Sundrum Compactification Scenario}'',
Phys. Rev. {\bf D60} (1999) 107505, hep-ph/9907218.

\bibitem{Verlinde}
E.~Verlinde and H.~Verlinde,
``{\it RG-Flow, Gravity, and the Cosmological Constant}'',
hep-th/9912018.


\end{thebibliography}
\end{document}